 \documentclass[twocolumn,english,aps,prb,superscriptaddress]{revtex4}
\usepackage[T1]{fontenc}
\usepackage[latin9]{inputenc}
\usepackage{amsmath}
\usepackage{amssymb}
\usepackage{graphicx}
\usepackage{esint}
\usepackage[english]{babel}
\usepackage{epsfig}

\def\beq{\begin{equation}}
\def\eeq{\end{equation}}
\def\beqa{\begin{eqnarray}}
\def\eeqa{\end{eqnarray}}

\def\e{\epsilon}

\def\D{\Delta}
\def\del{\delta}
\def\e{\epsilon}
\def\cH{{\mathcal H}}
\def\cL{{\mathcal L}}

\def\cdag{c^{\dagger}}

\def\del{\delta}

\newcommand{\bra}[1]{| #1 \rangle}
\newcommand{\ket}[1]{\langle #1 |}

\begin{document}

\author{Yonatan Dubi}
\email{jdubi@bgu.ac.il}
\affiliation{Department of Chemistry and the Ilse Katz center for Nano-Science, Ben-Gurion University of the Negev, Beer-Sheva 84105, Israel}

\title{Origin of thermoelectric response fluctuations in single-molecule junctions}
\date{\today}


\begin{abstract}
 The thermoelectric response of molecular junctions exhibits large fluctuations, as observed in recent experiments [e.g. Malen J. A. {\sl et al.}, Nano Lett. {\bf 10}, 3406 (2009)]. These were attributed to fluctuations in the energy alignment between the highest occupied molecular orbital (HOMO) and the Fermi level at the 
electrodes.  By analyzing these fluctuations assuming resonant transport through the HOMO level, we demonstrate that fluctuations in the HOMO level alone cannot account for the observed fluctuations in the thermopower, and that the thermo-voltage distributions obtained using the most common method, the Non-equilibrium Green's function method, are qualitatively different than those observed experimentally. We argue that this inconsistency between the theory and experiment is due to the level broadening, which is inherently built-in to the method, and smears out any variations of the transmission on energy scales smaller than the level broadening. We show that although this smmearing only weakly affects the transmission, it has a large effect on the calculated thermopower.  Using the 
theory of open quantum systems we account for both the magnitude of the variations and the qualitative form of the distributions, and show that they arise not only from variations in the HOMO-Fermi level offset, but also from variations of the local density of states at the contact point between the molecule and the electrode.                                                                                                                                                         
\end{abstract}

\maketitle


\section{Introduction}
Improving the thermoelectric (TE) energy conversion efficiency relies on increasing the thermoelectric response (Seebeck coefficient) 
and reducing the thermal conductance of the device \cite{Nitzan2007,dubi2011colloquium}. Metal-single molecule-metal junctions ('molecular junctions') seem promising in both  aspects: their thermal 
conductance is small due to the mismatch between the vibrations of the electrodes and the molecule, and their TE response should be large due to the 
well-defined resonant structure of the electron transport (via the molecular HOMO or LUMO levels) \cite{Malen2010}. This observation, along with the notion that TE response 
can shed light on transport mechanisms in molecular junctions \cite{Paulsson2003,Koch2004,Segal2005},  have initiated in large number of experimental \cite
{Reddy2007,Baheti2008,Malen2009,Malen2009a,Tan2011,Yee2011,Widawsky2012} and theoretical studies \cite
{Murphy2008,dubi2008thermoelectric,Liu2009,Bergfield2009,Liu2009a,Wang2010,Leijnse2010,Sergueev2011a,Liu2011,Liu2011a,Quek2011,Stadler2011,Nikolic2012,Burkle2012} on TE effects in molecular junctions. 

Most notable are a series of impressive experiments in which TE conversion using a single-molecule junction was demonstrated \cite{Reddy2007,Baheti2008,Malen2009,Malen2009a,Malen2010}. In these experiments, a single molecule 
(usually Benzene rings with various end groups) is trapped between an Au substrate and an Au scanning tunneling microscope (STM) tip, which are held at a constant temperature difference $\D T$. A voltage bias $\D V$ (here called 'thermo-voltage') is then 
applied between the tip and substrate to reach a 
state of zero current flowing through the molecule. The Seebeck coefficient is defined as (minus) the slope of $ \D V (\D T)$ as a function of $\D T$ in the linear response regime (i.e. $\D T \rightarrow 0$). In the experiments the value of $ \D V$ is strongly fluctuating, and repeating the 
experiment many times results in a broad distribution of $\D V$,  as can be seen in Fig.~3 of Ref. \cite{Malen2009}
(and inset of Fig.~\ref{fig2}(b) here). Using the typical $\D V$ (where the voltage distribution displays a maximum), the 
authors of Ref.\cite{Reddy2007,Malen2009} find typical Seebeck coefficients of the order of $\sim 10 \mu$V/K for single molecule junctions. 

The observed variations in the thermoelectric response do not seem to be simple (Gaussian) noise, and the distributions of the thermo-voltage $\D V$ may have a 
well-defined double-peak structure \cite{Reddy2007,Malen2009}. The authors of Ref.~\onlinecite{Malen2009} concluded that the fluctuations (also reported in 
thermoelectric measurements of metal-fullerene-metal junctions \cite{Yee2011}) are caused by variations of the position of the HOMO level, which are translated into  
variations in the energy offset between the HOMO level and the Fermi energy of the contacts, $E=E_{HOMO}-E_F$, as the junction reconstructs. Assuming that the transport is dominated by a single resonant level and using the Landauer 
formalism (or Green's function method) for thermoelectric transport (see, e.g. Refs. \onlinecite{DiVentra2008,dubi2011colloquium} ) they quantified the variations $\del E$ in $E$ and found them to be $\del E \sim 2.5 eV$, of the order of the average offset $E$ itself.

In this paper we argue that besides the variations in the offset energy between the HOMO level and the Fermi energy, there is an additional contribution to the 
fluctuations in the thermopower, which is the strong variations of the LDOS (as a function of energy) at the point of contact between the molecule and the 
electrodes. We use a toy model to verify this effect. The main results are (I) Within the Landauer or Green's function formalism for thermoelectric transport, any 
variations in the transmission function which are smaller than the self-energy (or level broadening) are suppressed. The result is a smooth transmission function and 
a smooth thermopower (as a function of energy).(II) The thermopower is extremely sensitive to small fluctuations of in the transmission function. As a result, the suppression of fluctuations in the transmission function, inherent to the Green's function method, strongly affects on the calculated thermopower, and results in an over-estimation for the 
fluctuations in the HOMO level-Fermi level energy offset fluctuations. (III) The thermo-voltage histogram obtained using the Green's function method is qualitatively 
different from the histograms obtained experimentally, and (IV) all of the above draw-backs of the Green's function formalism can be overcome by using the open quantum system approach to calculating the thermo-electric response in molecular junctions. 

The paper is organized as follows. In Sec.~\ref{sec1} we derive the estimation for the fluctuations in the HOMO-Fermi level energy offset from the experimental values 
of thermopower for typical molecular junctions, using a simple description of the molecular junction as a resonant level and using the Landauer formula. In Sec.~\ref{sec2} we present the open quantum systems approach, and using a simple model for the molecular junction we show that this approach for calculation of thermopower 
yields results which are in qualitative agreement with the experimental results. In Sec.~\ref{sec3} we show that, {\sl for the same model}, the Green's function approach 
gives very different results, due to the inherent suppression of fluctuations built in to the method (due to the presence of a self-energy of level broadening). We demonstrate that the thermopower is very sensitive to fluctuations, and thus the suppression of (small) fluctuations in the transmission strongly affects the calculation of thermo-power. Sec.~\ref{sec4} is devoted to summary and discussion.

\section{Analysis of thermopower fluctuations using the Landauer formula for a resonant level} \label{sec1}
As a starting point, we consider the data presented in Fig.~3 of Ref.~\onlinecite{Malen2009} (also inset of Fig.~\ref{fig2} here), where the thermo-voltage 
histograms of a  2',5'-dimethyl-4,4''-tribenzenedithiol (DMTBDT) molecular junction are 
presented. We observe several notable features: (i) Noise is present in the measurement: even at $\D T \approx 0$ there is a finite thermo-voltage distribution,  with an average signal offset of a few tens of microvolts, (ii) the typical form of the thermo-voltage distribution seems Gaussian, (iii) a double-peak structure of the distribution emerges, indicating two dominant values of the Seebeck 
coefficient, $S_1\sim 4.6 \mu$V/K and $S_2\sim 15 \mu$V/K.

 In order to estimate the variation in the molecular HOMO required to obtain these values, we use the standard 
Landauer formalism . Within this formalism, the conductance is given by $G=\frac{2e^2}{h} \tau(E_F)=G_0 \tau(E_F)$, and the Seebeck coefficient can be approximated as \beq 
S\approx -\frac{\pi^2 k^2_B T}{3 e} 
 \frac{\partial \log(\tau(\e))}{\partial \e} \left.  \right|_{\e=E_f} ~, \label{LandauerS} \eeq where $\e$ is the electron energy, $\tau(\e)$ is the electron transmission function, $e$ is the proton charge, $k_B$ the Boltzmann constant and $T$ is the average temperature. 
 
We proceed by assuming that the transport through  molecule is characterized by a transmission through a resonant level. To justify this, as well as the toy model considered 
in the next section, we briefly review the literature on theoretical calculations of thermopower in molecular junctions. The most common theoretical tool to calculate the 
transmission (and from it the thermopower) is the combination of non-equilibrium Green's function (NEGF)  and 
density-functional theory (DFT) \cite{DiVentra2008}, where the transmission function is evaluated using the Green's function as obtained from DFT. Within this 
approach, the transmission function is calculated as a function of gate voltage, and from it the thermopower is calculated, also as a function of gate voltage (see, e.g. Refs. 
\onlinecite{Liu2009,Stadler2011,Tan2011,Quek2011,Balachandran2012}). While in transport experiments gating of a molecule can be achieved 
\cite{Lee2003,Park2000,Li2006}, such a dependence of the thermopower on the gate voltage was never studied experimentally, and experiments always carry a statistical nature as discussed above. However, to our knowledge the 
statistical nature of thermopower in molecular junctions has never been addressed (although it has been addressed in the context of transmission \cite{Reuter2012} or thermopower of atomic wires \cite{Pauly2011}).  Nevertheless, to justify our model we observe that in the theoretical calculations, 
the transmission function seems always to resemble a resonant level, at least close to the HOMO or LUMO levels. \cite{Liu2009,Stadler2011,Tan2011,Quek2011,Balachandran2012}. Thus, considering a resonant level model is a common description even of realistic molecules with electron interactions.     

For a given resonant molecular level with energy $E_{HOMO}$ and contact Fermi energy $E_F$, the typical form of the transmission function is a resonant Lorentzian, 
$\tau(E)=\frac{\Gamma^2}{E^2+\Gamma^2}$, where $E=E_{HOMO}-E_F$ is the energy offset and ~$\Gamma$ is the contact-induced effective level broadening. The resulting  
Seebeck coefficient is of the form
\beq S\approx -\frac{\pi^2 k^2_B T}{3 e} \frac{2 E}{E^2+\Gamma^2}~~. \label{Seebeck}\eeq 

Noting the double-peak structure of the voltage distribution, we thus consider that there is a typical shift in the HOMO level $\del \e$. Taking the conductance value to be $G\sim 0.01 G_0$, one needs to solve simultaneously 
$
 \frac{\Gamma^2}{ E^2+\Gamma^2} = 0.01~,
 -\frac{\pi^2 k^2_B T}{3 e} \frac{2  E}{ E^2+\Gamma^2} = 4.6 \mu \textrm{V/K} ~, 
-\frac{\pi^2 k^2_B T}{3 e} \frac{2 ( E+\del \e)}{(E+\del \e)^2+\Gamma^2}= 15 \mu \textrm{V/K}~~. $ These equations yield (at ambient temperature) 
$\Gamma\sim 0.31$ eV, $ E \sim 3.1$eV and $\del \e \sim 2.2$ eV, such that $\del \e /  E \approx 0.7$, similar to the value obtained in Ref.~\onlinecite{Malen2009}. This means that the molecular level alignment with 
the contact Fermi level is changing by $\sim 2$ electron-volts at every reconstruction of the junction. This large variation (of the order of $ E$ itself) was attributed 
to variations in the junction contact geometry and intermolecular interactions. However, a recent study of transition voltage spectroscopy in 
molecular junctions (see Ref.~\onlinecite{Beebe2006} for description of the method) demonstrated that the variation of $ E$ are of the order of $\sim 1$eV, \cite{Guo2011} much smaller 
than the variation required to generate the value above. This discrepancy suggests that there is an additional factor contributing to the variations in thermopower. 

The discrepancy describes above hints that there is a flaw in the analysis of the thermopower fluctuations using the Landauer formula with the resonant level model. As we will 
show in the following sections, the flaw is that the Landauer formula inherently suppresses any variations in the transmission function (as a function of molecular level 
energy $E$) which are on an energy scale smaller than the level broadening (or electron self-energy), and since the thermopower is very sensitive to variations in the 
transmission, this suppression strongly affects the thermopower. Thus, fluctuations in $E$ sample a thermopower function $S(E)$ which is "too smooth", resulting in the over-estimate of $\del E$. 

\section{Open quantum system approach to thermopower}\label{sec2}

In order to identify the origin of variations in the thermopower, we study a toy model of a molecular junction. We use the method of open quantum systems (OQS)  
(which was described in detail in previous publications \cite{pershin2008effective,dubi2008thermoelectric}) to show that the thermopower $S(E)$ exhibits variations as a 
function of $E$ which are on the same energy scale as the variations in the LDOS at the molecule-lead point of contact, and show that the thermo-voltage histogram resulting in 
fluctuations in $E$ is qualitatively similar to the experimental ones. In the next section we will compare the result ontained from the OQS method to those obtained by using 
NEGF method. 
 
\begin{figure}
\vskip 0.5truecm
\includegraphics[width=8truecm]{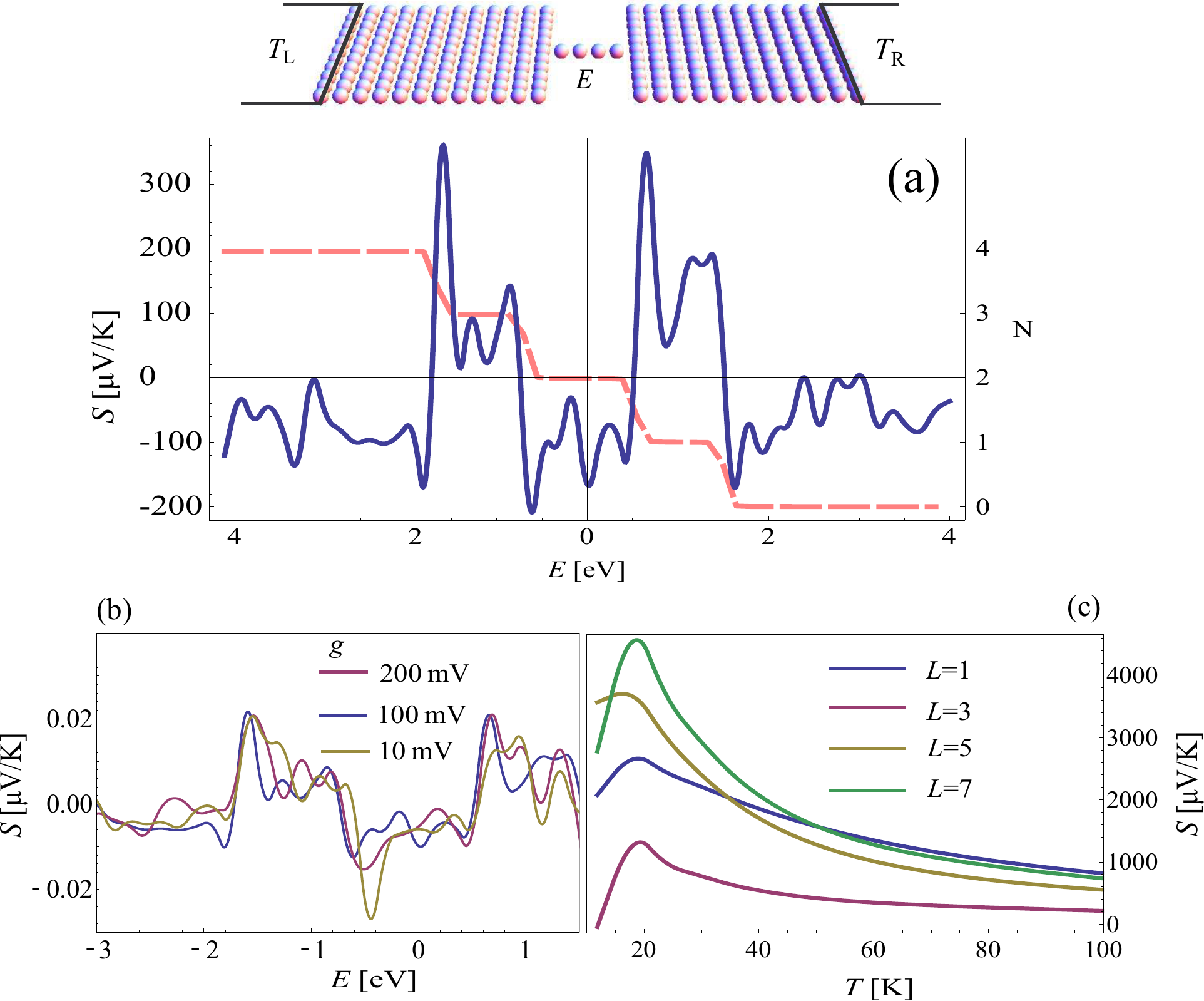}
\caption{Upper panel: schematic representation of the molecular junction. (a) Seebeck coefficient $S$ (solid blue line) and electron occupation of the molecular chain 
(dashed orange line) as a function energy $E$ (see text for numerical parameters). As the molecular levels cross the electrode Fermi energy (which is 
set as the zero energy) the occupancy of the dot change, and correspondingly the Seebeck coefficient changes sign. Oscillations of the Seebeck coefficients between the changes 
of sign are observed. (b) Same as (a) for different values of the electrode-molecule coupling, $200, 100$ and $10$ meV. The coupling has little effect on the Seebeck coefficient. (c) temperature dependence of the Seebeck coefficient (at $E=0$) for various molecular chain lengths, exhibiting a non-monotonic temperature dependence. \label{fig1} }
\end{figure}

Our toy model consists of two finite electrodes with a molecular bridge between them (upper panel of Fig.~\ref{fig1}), attached to reservoirs with different temperatures $T_L$
 and $T_R$ for the left and right edges respectively. The molecule is described by a simple chain with four atomic orbitals. The OQS method enables one to calculate  the 
electronic density in this non-equilibrium situation (a finite temperature difference between the electrodes). From the electronic density, one can calculate (via the Poisson 
equation) the voltage difference $\D V$ between the electrodes as a function of the temperature difference $\D T=T_R-T_L$. Repeating the calculation for different values of 
$\Delta T$ we then find a curve $\Delta V (\Delta T)$, and the linear slope is the thermopower (or Seebeck coefficient) $S=-\frac{\partial \D V}{\partial \D T}$. 

Essentially, OQS theory is a mapping of the system Hamiltonian $\cH$ (which in general includes all electronic degrees of freedom, electron interactions, junction geometry 
etc.) onto a master equation for the single-particle density matrix of the Lindblad form \cite{pershin2008effective}, 
\beq\dot{\rho} =
-i[\cH,\rho]+\cL [\rho]\label{Lindblad}~~.\eeq We consider a non-interacting tight-binding Hamiltonian of the form $ \cH=\cH_L+\cH_R+\cH_d+\cH_{c} $, where 
\beq \cH_{L,R}=-t \sum_{\langle
i,j \rangle \in L,R} \left(\cdag_i c_j + h.c.\right) \eeq are the tight-binding
Hamiltonians of the left and right leads respectively ($t$ is the hopping integral, which is taken as the unit energy, $t=1$eV from here on),
and \beq \cH_{d}=-t \sum_{\langle
i,j \rangle \in d}\left(\cdag_i c_j + h.c.\right) + \sum_{i \in d} (E-\mu)\cdag_i c_i \eeq is the Hamiltonian for the wire, which includes the usual hopping integral and an energy $E$ which can be tuned 
(for instance using a gate electrode experimentally). The energy is measured with respect to the electrodes' Fermi energy $\mu$ which is set as the zero energy ($\mu=0$). 

The coupling between the electrodes and the molecular wire is described by
 \beq\cH_{c}=
(g_L\cdag_{L}c_{d,0}+g_R\cdag_{R}c_{d,L_d}+h.c. )\eeq describes the
coupling between the left (right) lead to the wire, with
$\cdag_{L(R)}$ being the creation operator for an electron at the
point of contact between the left (right) lead and the wire, and
$c_{d,0}$ ($c_{d,L_d}$) destroys an electron at the left-most
(right-most) sites of the wire (we take here $g_L=g_R=g$). The external environment(s) are accounted for in Eq.~\ref{Lindblad} by the second term, which has the form 
\beq
\cL [\rho ]=-\frac{1}{2}(V^\dagger V\rho+\rho V^\dagger V)+V \rho V^\dagger
~~, \label{Lindbladian}\eeq where $V$ are the Lindblad $V$-operators which encode the properties of the environment, i.e. its temperatures and position (i.e. left or right electrode). An 
appropriate form for the $V$-operators is \cite{pershin2008effective} 
\beqa
V^{(L,R)}_{kk'}=\sqrt{\gamma^{(L,R)}_{kk'}f^{(L,R)}_D(\e_k)} \bra{k}
\ket{k'} \label{opeq}~~,\eeqa where $f^{(L,R)}_D(\e_k)=1 /\left(
\exp \left (\frac{\e_k-\mu}{k_BT_{L,R}}\right)+1 \right)$ are the
Fermi distributions of the left and right leads (with the corresponding temperature), $\mu$ the
chemical potential, and $\gamma^{(L,R)}_{kk'}$ are the overlap integrals between the $k$ and $k'$ states on the left (L) and right (R) edges of the electrodes. This form for the $V$-operators guarantees 
that at equilibrium (i.e. $T_L=T_R$) the diagonal elements of the density matrix are described by a Fermi function. 

We note that the Lindblad equation assumes a Markov approximation for the environment, 
which is reasonable since we are considering a system at room temperature (where quantum memory effect of the environment should not be important), and since we are interested in the steady state (and not 
the dynamics).  The numerical calculation is performed in the following way: (i) From the tight-binding Hamiltonian  the single-particle states and energies are calculated, and the $V$-operators are 
constructed according to Eq.~(\ref{opeq}) for different electrode temperatures $T_L$ and $T_R$. (ii) The $V$-operators are inserted to Eq.~(\ref{Lindblad}) and Eq.~(\ref{Lindbladian}), which are then solved in the steady-state (i.e. for $\dot{\rho}=0$). (iii) 
From the diagonal elements steady-state solution for $\rho$ and the wave functions, the electron density is calculated. (iv) By solving the Poisson equation, the voltage at the center of the electrodes is 
calculated. The voltage difference between the electrodes is the thermo-voltage, as it is induced by the temperature difference. the slope of the thermo-voltage as a function of the temperature difference 
is the thermopower. 

In Fig.~\ref{fig1} we plot the the Seebeck coefficient (solid blue line) and the electron occupation of the molecular chain 
(dashed orange line) of a model molecular junction as a function of a gate potential applied to the junction, i.e. as a function of the molecular energy level with respect to the Fermi level (which is set as the zero energy). The junction is composed of a series of four atomic orbitals connected to square two dimensional 
electrodes (upper panel). Numerical parameters are: electrode size $L^2=25 \times 25$, tight-binding hopping integral $t=1$eV (corresponding to a band-width of $W=8$eV), electrode-molecule coupling is $g=0.1$eV, Temperature is room temperature, $T=300K$ and the electrodes are kept at half filling. 

The first thing to be noted is that whenever a molecular orbital crosses the Fermi energy, the molecular occupation changes by one, and the thermopower changes sign. This is in accord with the known results from the Landauer formula for thermopower, and is due to the change from electron-dominated to hole-dominated transport every time the Fermi level is crossed (which also corresponds to a transmission resonance). The second important feature is that, in contrast to the result expected from the regular Landauer formula \cite{Paulsson2003,dubi2011colloquium}, the Seebeck coefficient exhibits strong variations with gate voltage. 

Before we proceed to discussing the origin of these variations, it is useful to compare the properties of the Seebeck coefficient as obtained from the OQS method to those 
known from the Landauer formalism. In Fig.~\ref{fig1}(b) the Seebeck coefficient as a function of gate voltage (same as in Fig.~\ref{fig1}(a)) is plotted for different values 
of the coupling between the electrode and the molecule, $g=200, 100$ and $10$meV. As expected from the Landauer formalism \cite{Quek2011}, there is little effect 
to the coupling on the magnitude of the Seebeck coefficient. In Fig.~\ref{fig1}(c) the Seebeck coefficient is plotted as a function of temperature for various molecular chain 
lengths. An inhomogeneous temperature dependence is found, originating from a crossover from coherent to incoherent transport, again in agreement with results obtained from the 
Landauer formalism (e.g. \cite{Segal2005}). These results demonstrate that the main physical features of the thermopower which are present in the Landauer formalism, also 
appear within the framework of the OQS theory.

\begin{figure}[h!]
\vskip 0.5truecm
\includegraphics[width=8.5truecm]{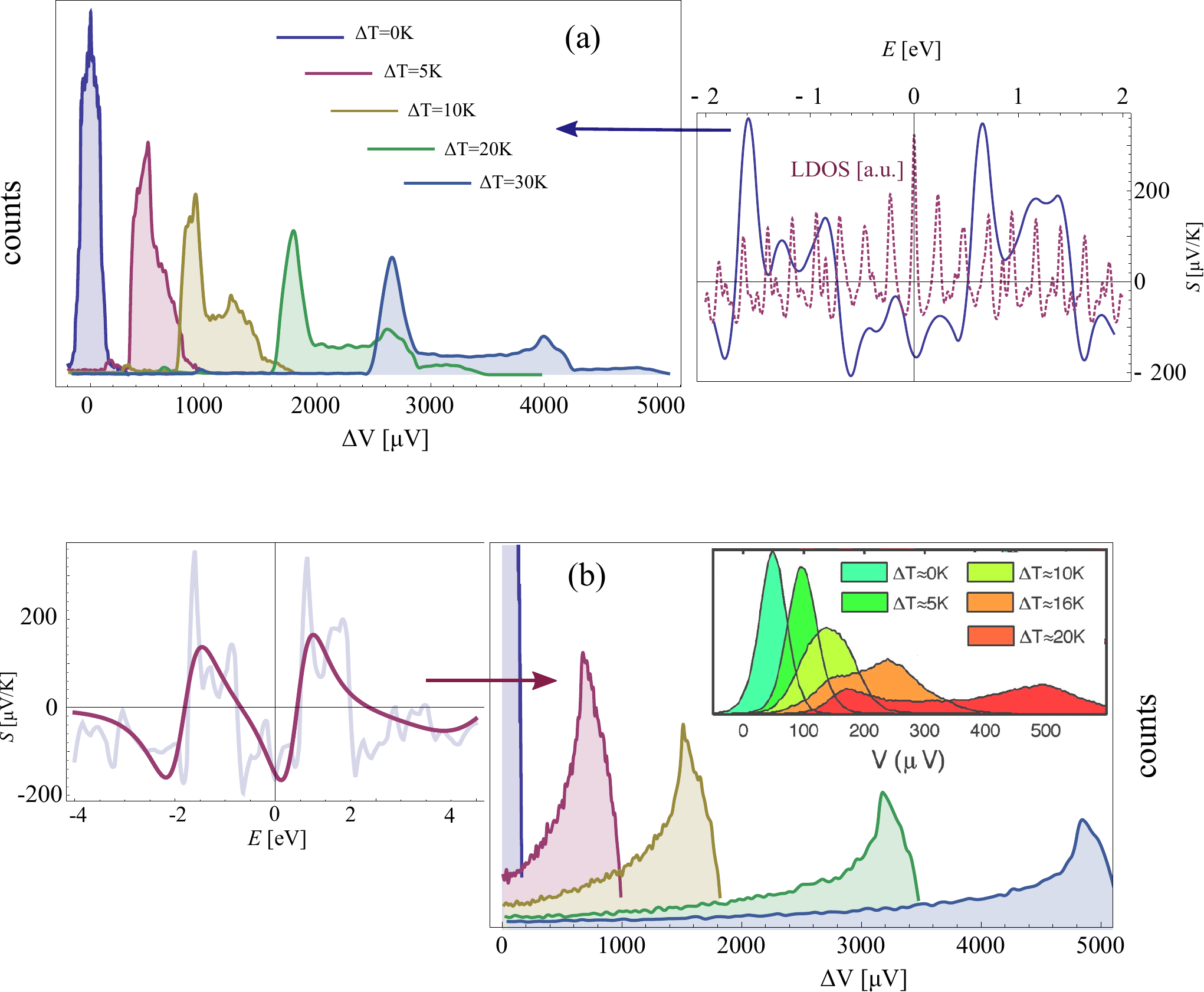}
\caption{Thermo-voltage $\D V$ histograms for different temperature differences $\D T=0,5,10,20,30$K, obtained by taking $\D V=S(E) \D T$ with $E$ normally distributed around $E=0.3$eV with width $\Gamma_E=0.2$eV. (a) Histograms obtained from the thermopower $S(E)$ obtained using OQS theory (right panel solid line). The LDOS at the point of contact between the molecular chain and the electrode is also plotted (right panel solid line). (b) Same as (a), but for the Seebeck coefficient $S(E)$ obtained using the Landauer formula, which does not exhibit strong local fluctuations. Inset: Experimental histograms of $\D V$ for a DMTBDT molecular junctions, taken with permission from Ref.~\cite{Malen2009}.\label{fig2}}
\end{figure}

We now turn to calculating the distribution of the thermo-voltage $\D V$ across the molecular junction. To obtain the distribution, we calculate the temperature-difference 
induced voltage $\D V=S(E) \D T$, taking the gate voltage of the molecule (i.e. the HOMO-Fermi energy offset) $ E$ to be a random variable, normally distributed around $ 
E_0=0.3$eV with a width $\Gamma_E=0.2$eV. While these values are rather tdifferent then experimental values (probably $E_0$ is bigger in experiments) we point that we are aiming at {\sl qualitative} 
similarity to experiment, to point the origin of the variations, rather than to analyze realistic junctions. To mimic the experiment, we also apply a small variation (two degrees Kelvin) to the electrode 
temperatures (note that in the experiments, a finite thermo-voltage distibution was observed even at $\D T \approx 0$,see inset of Fig.~\ref{fig2}(b), indicating the existence of a small temperature 
difference, probably due to Johnson noise \cite{Malen}. This does not affect our results or conclusions). 

In Fig.~\ref{fig2}(a) the distribution of thermo-voltage is plotted for temperature differences $\D T=0, 5, 10, 20, 30$K. The distributions qualitatively resemble the 
experimental distributions, exhibiting a broad double-peak structure. To understand the origin of the distribution shapes, in the right panel of Fig.~\ref{fig2}(a) the Seebeck 
coefficient is again plotted (solid 
line). Due to the strong sensitivity of $S$ on $E$, even a relatively small variation in $ E$ (of the order of $\Gamma_E\sim 0.2$eV) can include several 
maxima of $S$, giving rise to the different peaks in the distributions. 

To understand the origin of the sensitivity of $S$, in the right panel of Fig.~\ref{fig2}(a) we  
plot the local density of states at the point of contact between the electrode and the molecular chain (dashed line). One can see that the 
variations of $S(E)$ and the LDOS vary on the same energy scale. Note that these variations are a surface effect, 
and are not due to the electrode level spacing (the average level spacing is $\sim 0.01 eV$). We thus conclude that it is 
the variations in the LDOS (which are on an energy scale much smaller than the energy difference between the molecular levels) that give rise to the sensitivity of $S$ (we note that the total DOS of the electrodes is a much smoother function with no observed variations on these energy scales). We point that the LDOS variations are not 
a finite-size effect, but rather are a surface effect. Calculating the LDOS to systems as large as $200 \times 200$ (where the level spacing is $\sim 0.0005$eV), we found that local variations in the LDOS persist with roughly the same energy scale of $\sim 0.1$eV (although their magnitude somewhat changes). 

On the other hand, when considering the thermopower as obtained by the Landauer formula, the local variations in the LDOS are not taken into account, or rather they are smeared 
by the electron self energy which is reflected through the level broadening $\Gamma$ in Eq.~(\ref{Seebeck}). The resulting $S(E)$ is a much 
smoother function. As a first demonstration of this effect, here we use the formula for transport and thermopower through a resonant level, Eq.~\ref{LandauerS}. The positions of the levels and the level 
broadenings are obtained by fitting the data of Fig.~\ref{fig1}(a).  In the right panel of Fig.~\ref{fig2}(b) $S(E)$ is plotted using Eq.~(\ref{Seebeck}) and the parameters of the junction (i.e. resonances 
and widths fitted to Fig.~\ref{fig1}(a)). For comparison, $S(E)$ obtained from OQS theory is plotted as a bright line in the background. The resulting $\D V$ distributions are plotted in 
Fig.~\ref{fig2}(b) for different $\D T$. The distributions obtained using Eq.~\ref{Seebeck} are quite different from the experimental results (inset of 
Fig.~\ref{fig2}(b)), both in terms of shape and in the lack of the double-peak structure. 

\section{Green's function analysis of the toy model}\label{sec3}
Let us now examine the same model within the Green's function formalism. In this formalism, and for our non-interacting toy model, the transmission function $\tau (E)$ is given by \cite{Meir1992,DiVentra2008} 
\beq
\tau (E) = {\mathrm Tr} \left( \Gamma_L G^r(E) \Gamma_R G^a(E) \right) \label{tau_E} \eeq 
where $G^{r,a}(E)=\left( E-\cH \mp i \Sigma \right)$ are the retarded and advanced Green's functions, and $\Gamma_{L,R}$ represent the level broadening due to the electrodes, typically a few hundreds meVs \cite{Bergfield2009,Burkle2012}. In the basis of atomic orbitals (as the Hamiltonian is written) $\Gamma_{L,R}$ are diagonal matrices, with $\Gamma_{L,R}(n,n)=\Gamma$ if $n$ is in the left or right edges of the electrodes, and zero otherwise. We take $\Gamma=0.3$eV in the numerical example below. $\Sigma$ is the self-energy, which in the non-interacting case is only due to the electrodes, and hence $\Sigma=\Gamma_L+\Gamma_R$. 

\begin{figure}
\vskip 0.5truecm
\includegraphics[width=8.5truecm]{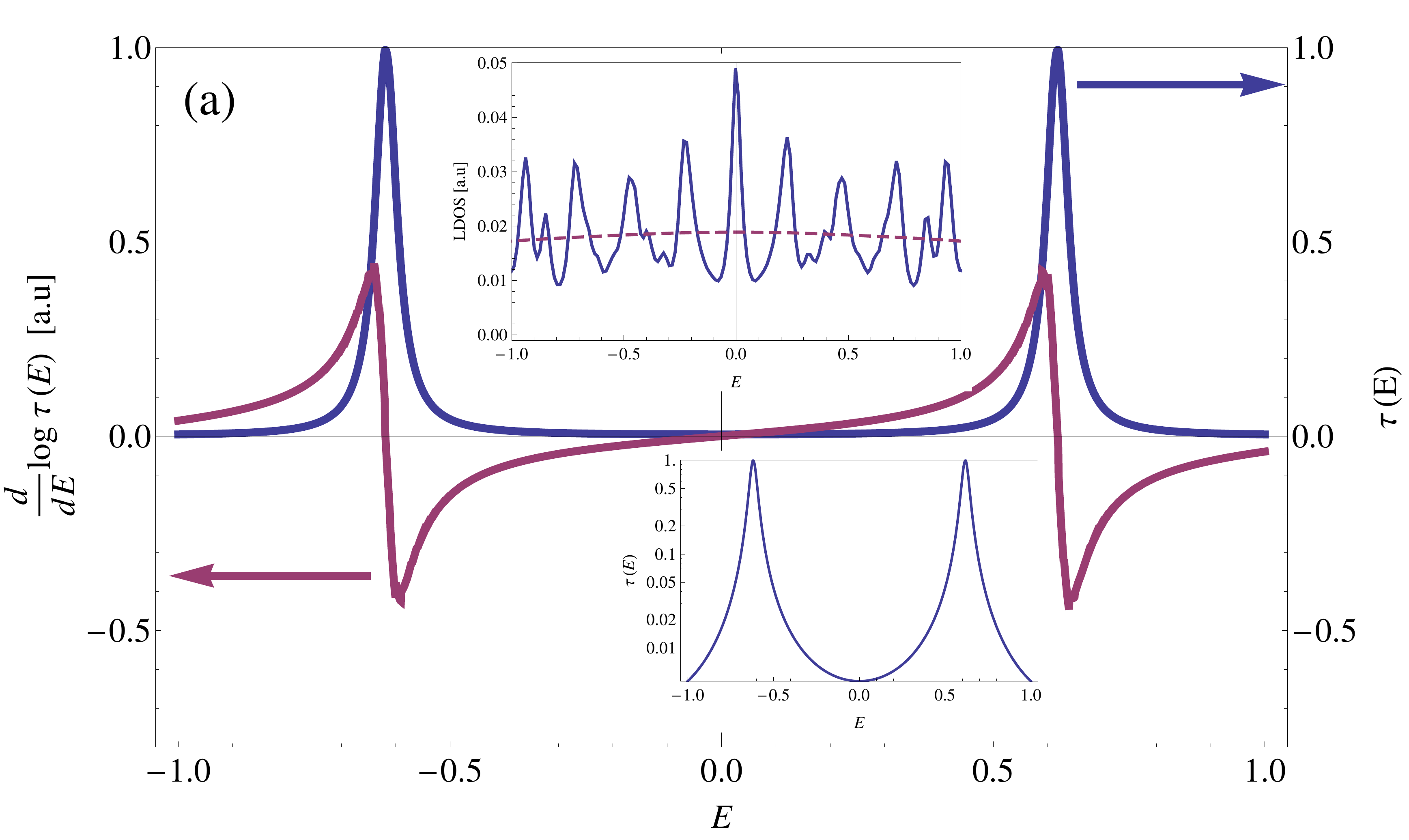}
\includegraphics[width=8.5truecm]{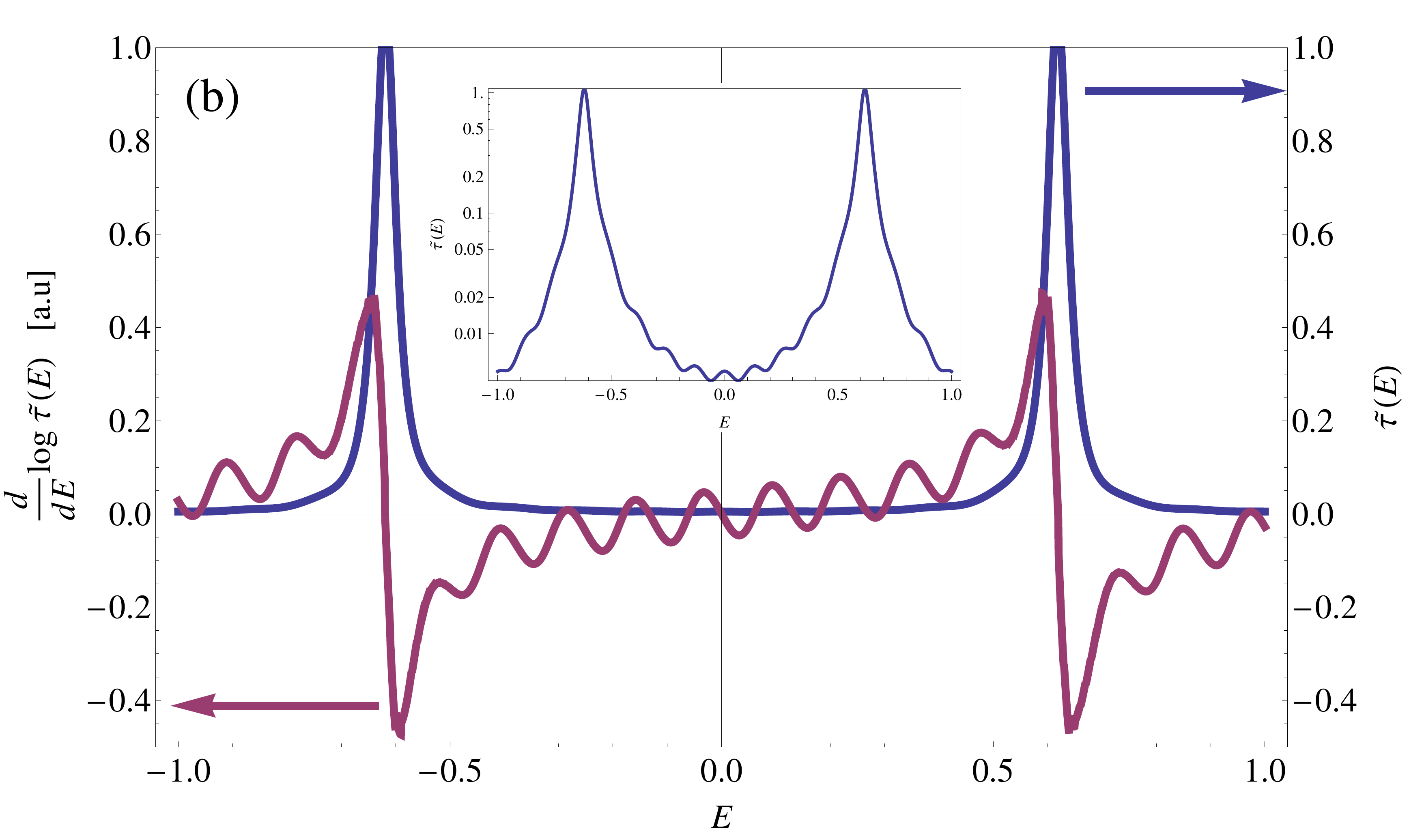}
\caption{(a) Transmission function $\tau(E)$ and thermopower $S(E)$ as a function of energy offset for the tight-binding toy model. The transmission and the thermopower are perfectly smooth and disregard any variations due to the DOS. Upper inset: the local density of states at the point of contact between the wire and the electrode (sloid line), and the same LDOS broadened by $\Gamma$ (dashed line). (b) same as (a) for the artificial transmission function $\tilde{\tau}(E)=\tau (E) \left( 1+0.1 \cos (E/E_0\right)$. The oscillations in the transmission function are hardly visible, yet they induce large variations in the thermopower $S(E)$. Inset: $\tilde{\tau}(E)$ on a log-scale, only there the oscillations are visible.   }
\label{fig4}
\end{figure}

Once the transmission is calculated, the thermopower can be calculated directly using Eq.~(\ref{LandauerS}).  In Fig.~\ref{fig4}(a) the transmission function $\tau (E)$ and the its logarithmic derivative (proportional to the thermopower $S(E)$) of the toy model are 
plotted (as a function of the gate energy). For comparison, the local density of states at the point of contact between the wire and the electrode is plotted in the upper inset (the dashed line in the inset is the DOS when broadened by $\Gamma=0.3$eV). As seen, the oscillations in the LDOS are completely smeared in the transmission 
(plotted as a function of the gate energy). For comparison, the local density of states is plotted in the upper inset (the dashed line in the inset is the DOS when broadened by $\Gamma=0.3$eV). As seen, the oscillations in the LDOS are completely smeared in the transmission 
function and hence in the thermopower. The lower inset shows the transmission function on a log scale, verifying that the variations are really smeared out. To demonstrate that this smearing has a large 
effect on the thermopower, we consider an artificial transmission function $\tilde{\tau}(E)=\tau (E) \left( 1+0.1 \cos (E/E_0\right)$ with $E_0=0.02$eV. This transmission exhibits oscillations on an energy 
scale $E_0$. However, they are chosen in such a way that they are hardly visible if the transmission is observed in its full scale, as seen from Fig.~\ref{fig4}(b), where $\tilde{\tau}(E)$ is plotted. In 
the inset the transmission is plotted on a log scale, and only there the oscillations are visible. On the other hand, the thermopower exhibits very strong oscillations of considerable size even if the 
oscillations are hardly observed in the transmission functions. This is due to the extreme sensitivity of $S(E)$ to local variations of $\tau(E)$, reflecting its derivative structure. 

\section{Summary and discussion}\label{sec4}
In summary, we have analyzed the variations in the thermoelectric response of a metal-single molecule-metal junctions, based on detailed examination of experimental results. 
The experimental results, namely the width of the thermo-voltage variations and the shape of the thermo-voltage distributions, cannot be accounted for by 
only assuming variations in the misalignment between the molecular HOMO and the electrode Fermi energy. Using the theory of open quantum systems we qualitatively reproduce the 
experimental results, and show that they may originate in a combination of the level misalignment variations and variations of the local density of states at the point of contact between the molecule and the electrode, specifically the STM tip in typical molecular junction experiments. 

To put it differently, we found that in order to explain the 
thermopower variations in molecular junctions the electronic transmission function cannot have a simple Lorentzian form, but rather should have a more complicated form which 
includes variations on a scale smaller than the HOMO-LUMO gap. Such variations may be induced by the variations in the LDOS at the tip-molecule point of contact, originating 
from the STM tip structure, impurities or trapped states, etc. The use of the Green' function formula smears out all variations on energy scale smaller than the level 
broadening (typically a few hundred meVs). This smearing has a very small effect on the transmission and hence the conductance of the molecular junction, but have a strong 
effect on the thermopower. Thus, caution needs to be taken when using the NEGF method for calculation of themopower. We note a recent paper \cite{Evans2009} arguing that the 
NEGF method has limitations even when calculating the conductance of a molecular junction (although the argument is different then presented in this paper). 
 
The effects of STM tip structure on local transport measurements have been extensively discussed in the STM literature \cite
{Tersoff1990,Pelz1991,Hofer2003,Passoni2009,Kwapinski2010}, and were recognized to have an important contribution to the overall charge transport, via the tip LDOS. 
Our results imply that a similar importance of STM tip structure appears in thermo-electric measurements. Consequently, a direct extraction of relevant parameters such as the 
HOMO-Fermi level offset from thermo-electric measurements requires taking the electronic properties of the metal-molecule interface into account.  

The author wishes to thank J. Malen for valuable comments on the manuscript.

\end{document}